# GenAI Integration into Engineering Education: A Case Study of an Introductory Undergraduate Engineering Course

[a]Kadir Kozan, [b]Ozgur Keles, [a]Sihan Jian, [c]Serkan Ayvaz, [c]Krzysztof Sierszecki, & [a]Sewon Joo

[a]Florida State University, Department of Educational Psychology and Learning Systems, Tallahassee, FL, USA

[b]University of North Carolina at Charlotte, Department of Mechanical Engineering and Engineering Science, Charlotte, NC, USA

[c]University of Southern Denmark, The Maersk Mc-Kinney Moller Institute, Odense, Denmark

**Abstract**

GenAI has a potential to enhance the learning and teaching processes in engineering education. For instance, GenAI feedback on students' task performance can be effective depending on when such feedback is provided. However, little is known about how engineering faculty and instructors discover such potential within the scope of their instruction when they try out the technology for the first time. To this end, this study purported to describe an engineering instructor's and seven teaching assistants' initial experiences of integrating GenAI into their undergraduate engineering course and the corresponding changes in students' formative exercise performance. An embedded descriptive single case study design was employed. The corresponding research data included four interviews conducted at the beginning, middle and end of an academic semester, and students' formative exercise performance. Overall, after GenAI integration, students' formative exercise performance increased, and a critical and reflective practice of learning about how to integrate GenAI into instruction provided informative insights. Still, technology integration stayed at the level of replacing other instructional methods or increasing the efficiency of solving coding problems. It turned out to be exciting and surprising for students to be able to use GenAI in course work even though their use of the technology weakened over time. Our findings suggest that engineering teaching staff's initial experimental experiences with GenAI integration can be informative and provide context-specific practical insights. Therefore, it is reasonable for higher education institutions to encourage such experiences especially when there is a lot of unknown regarding an emerging technology.

**1. INTRODUCTION**

Generative artificial intelligence (GenAI) integration into engineering education (EE) has been subject to practice and emerging research referring to positive outcomes (e.g., Birtek et al., 2023; Fan et al., 2023) and potential limitations (e.g., Ndiaye et al., 2023; Pham et al., 2023; Qadir, 2023). Qadir (2023) highlighted that even though GenAI is both promising and defective, there will be more sophisticated technology coming up for which we need to be ready by understanding how EE can adapt to technological changes. Therefore, we need more conclusive insights into GenAI effects on learning and teaching in EE in the classroom (e.g., Qadir (2025). One valuable source of such insights are undergraduate course instructors who design and deliver engineering instruction and integrate technology into it. Accordingly, it is reasonable to research engineering instructors' GenAI integration efforts to understand how the integration works thus informing the design and development of GenAI-enhanced EE.

Further, this paper aims to distinguish between: (a) AI in education broadly, including intelligent tutoring systems and adaptive learning; and (b) AI-assisted coding education, where GenAI serves as a programming assistant, tutor, or collaborator, which is recent and substantially underexamined. While short-term gains (e.g., faster task completion) may occur, potential long-term drawbacks, such as reduced skill development, require longitudinal analysis (Ndiaye et al., 2023). Thus, this study's purpose is to understand how intentional and exploratory GenAI integration into introductory undergraduate engineering courses would work and enhance students' performance across a full semester. To this end, this study aimed to gain insights into an engineering faculty's and teaching assistants' initial GenAI integration and their thought processes. Such insights can significantly inform our understanding of how to integrate GenAI into EE more effectively.

Despite previous attempts to integrate GenAI into EE and to understand how it works and enhances student learning outcomes (e.g., (e.g., Ho and Lee, 2023; Lee & Low, 2024; Uddin et al., 2024), there is limited research on GenAI integration in the engineering classroom (Qadir, 2025). Thus, there is still a need for applied research on GenAI in higher education (Kurtz et al., 2024) including software EE (Nguyen-Duc et al., 2025), and research that qualitatively examines the initial processes engineering teaching staff go through while trying to figure out how to integrate GenAI. To this end, this study addressed the following questions:

- What changes in students' formative practice exercise performance occurred in an introductory engineering course where the course instructor and teaching assistants intentionally explored GenAI integration for the first time?
- What did the instructor and teaching assistants describe as components of the GenAI integration experience, and how did their descriptions align with Hughes's (2005) technology integration levels?

## 2. LITERATURE REVIEW

### 2.1. Generative Artificial Intelligence and Engineering Education

Artificial intelligence integration in EE started with basic automated instructional systems and gradually developed towards advanced GenAI tools like ChatGPT, GitHub Copilot, and Google's Bard, substantially enhancing curriculum design and instructional methods (Mustapha et al., 2024; Randall et al., 2024; Sengul et al., 2024). The rapid emergence of these tools has made them important for EE, significantly influencing instructional design and student interactions (Lee & Low, 2024; Yelamarthi et al., 2024).

GenAI integration into EE can occur in various ways including creating questions or assignments and lesson plans (Johri et al., 2023), and previous research on integrating GenAI into EE produced promising insights, limitations, boundary conditions, and concerns. For instance, Uddin et al. (2024) found that integrating ChatGPT into civil engineering coursework led to more detailed and accurate responses by students, evidencing enhanced comprehension and retention. However, GenAI showed limited capacity while trying to solve chemical engineering problems at higher levels of Bloom's taxonomy including creativity and evaluation (Shahid & Walmsley, 2026). Questions about misinformation and overreliance have also shown up as potential limitations (e.g., Bravo & Cruz-Bohorquez, 2024; Qadir, 2023). Accordingly, providing effective guidelines and targeted training affected the quality of GenAI integration into EE (e.g., Walter, 2024).

GenAI's adaptive learning capabilities can improve the personalization of educational experiences (Alasadi & Baiz, 2023; Chen et al., 2020; Li et al., 2024). For instance, Qadir (2023) stated that LLMs can complement human instruction in EE by enhancing personalized learning, tutoring, and idea generation, although concerns about misinformation and overreliance persist thereby leading to trust issues (Goldshtein et al., 2025). Additionally, platforms such as Khan Academy's GenAI-powered Khanmigo offers various activities for learners to choose, provides adaptive feedback as well as meaningful and advanced output through authentic activities (Shetye, 2024).

GenAI has been shaping learning contexts as well given its capacity to efficiently create educational content and activities (Mollick & Mollick 2023; Nikolic et al., 2023). GenAI-enhanced virtual laboratories, for instance, have expanded the scope of laboratory activities by encouraging student engagement in experiments typically constrained by safety, logistical, or financial limitations, and by implementing personalized and adaptive learning through authentic activities (e.g., Ho & Lee, 2023; Murali et al., 2024; Tuyboyov et al., 2025). GenAI can also help create EE learning experiences using other technological platforms. For instance, Ho and Lee (2023) reported a successful case of designing and developing learning experiences in the Roblox metaverse by using ChatGPT that increased students' engagement and understanding.

From a curriculum and content perspective, GenAI has proven useful for updating curricula to match rapid industry advancements and emerging technologies (Mustapha et al., 2024; Lee & Low, 2024). For instance, GenAI tools have successfully been employed in mechanical EE to create dynamic simulations, design scenarios, and coding exercises, enhancing students' ability to practically apply theoretical knowledge (e.g., Mustapha et al., 2024). In software EE, conversational GenAI improved learning outcomes by automating routine programming tasks, thus enhancing students' engagement with complex coding assignments (Randall et al., 2024).

The benefits of GenAI come with challenges and concerns, particularly around academic integrity and ethics. Concerns about GenAI-assisted plagiarism are notable as highlighted by studies demonstrating the challenges of detecting the misuse of GenAI-generated content through conventional plagiarism methods (e.g., Dwivedi et al., 2023; Gao et al., 2022; Zhou & Wang, 2024). The difficulty in reliably distinguishing GenAI-generated from human-generated code has prompted recommendations for educators to explicitly teach ethical use and critical assessment skills. For instance, rethinking assessments to adapt to these new realities is necessary (Gambhir et al., 2024; Nikolic et al., 2023). Unsurprisingly, engineering faculty expressed concerns about plagiarism and misinformation (Guillen-Yparrea & Hernandez-Rodríguez, 2024).

Ethical considerations such as algorithmic biases, privacy, and equitable AI access remain substantial issues too. Research highlights the need for institutional guidelines addressing these ethical concerns (e.g., Lee & Low, 2024). Technical barriers, including high implementation costs and infrastructure demands, represent additional significant obstacles for effective GenAI integration (Alasadi & Baiz, 2023; Synekop et al., 2024). Accordingly, Mozgovoy et al. (2023) called for equipping students and faculty with GenAI literacy to use these tools effectively and to critically assess their limitations and potential biases, and all these challenges and opportunities require engineering educators update their skills and knowledge (Johri et al., 2023).

Overall, engineering faculty's or teaching staff's viewpoints are important for successful GenAI integration (Hassani et al., 2025; Simelane & Kittur, 2024). Even though engineering faculty acknowledge positive GenAI effects (Joskowicz & Slomovitz, 2024), they also have concerns including overdependence and ethics (Al Badi et al., 2024; Simelane & Kittur, 2024), and assessment disruption (Gambhir et al., 2024). Simelane and Kittur (2024) highlighted that engineering faculty tend to use ChatGPT as an ancillary tool and accept its advantages, but also have such concerns as students' overdependency. Engineering faculty with more practice-oriented roles are more open to GenAI integration while faculty with more education- or research-related roles have mixed support (Harper, 2024). Still, engineering teaching staff think that GenAI is crucial for EE (Al Badi et al., 2024).

**2.1.2. GenAI, and Instructional Effectiveness and Efficiency**

Conversational GenAI integration into software engineering courses has improved instructional effectiveness. For example, the use of GitHub Copilot for professional developers resulted in over 55% improvements in coding productivity (Peng et al., 2023). Similarly, studies indicated that GenAI-driven teaching assistants significantly enhance the learning experiences in solar energy engineering by guiding students through design processes, resulting in improved comprehension and performance (Sung et al., 2024). Such effectiveness seems to hold true for more general GenAI as well: Synekop et al. (2024) claimed that educators view ChatGPT as a potentially useful tool for enhancing specific instructional areas such as research skills, though its broader impact on instructional workload and pedagogical quality remains questionable. Given that teaching quality can be a strong contributor to faculty's integration of new technologies (Salajan et al., 2015), we still need to know how engineering faculty develop their sense of teaching quality through GenAI integration.

GenAI can also help improve complex engineering task performance. Liu and Yang (2024) integrated large language models (LLMs) into a system modeling and simulation engineering course to assist students with MATLAB programming, interdisciplinary concept understanding, and problem-solving. The implementation improved students' academic performance and notably reduced low-performing scores, demonstrating considerable effectiveness in enhancing student achievement and confidence in complex engineering tasks. Through structured guidelines and educational initiatives, Walter (2024) observed increased student awareness and engagement with GenAI tools. This study also identified that clear guidelines and dedicated training significantly improved student confidence and ethical GenAI use, although ongoing challenges like ensuring guideline adherence and accurately assessing the authenticity of student work continued to exist. Likewise, Bravo and Cruz-Bohorquez (2024) examined chatbots in electronic and mechatronic engineering instruction and revealed that students found chatbot assistance valuable in improving their coding skills, enhancing confidence in programming tasks, and increasing overall engagement.

Likewise, while exploring how ChatGPT influences mechanical engineering students' designs, Zhang et al. (2025) found median absolute deviation in the number of concepts to be 1.0 for ChatGPT-assisted teams versus 0.5 for those using traditional resources, reflecting greater exploration. In Gudoniene et al. (2023), all the participating students produced wireframes matching the intended design in the GenAI-assisted scenario whereas only 60% of those with a manual approach achieved it. Further, more students in the GenAI-assisted scenario completed the task faster. These positive effects of GenAI on instructional effectiveness and efficiency is not surprising given its potential to conduct professional engineering tasks: Lesage et al. (2024), for example, found that GPT-3 successfully generated a lab report's discussion section from student notes with an advanced readability score and no linguistic errors in mechanical EE.

In sum, GenAI integration into EE can streamline repetitive tasks, improve instructional efficiency, enhance task completion and student satisfaction, and promote educational scalability and resource availability. Research also revealed that engineering teaching staff are positive about GenAI integration into EE but also have some concerns. Yet, we still need more specific insights into: (a) engineering teaching staff's GenAI integration experiences by trying it out; and (b) GenAI integration level that can be achieved, and the changes that can be observed in student performance during such experiences.

## 3. THEORETICAL FRAMEWORK

We approached GenAI integration into EE through the lens of Hughes's (2005) three technology integration levels: (a) replacement; (b) amplification; and (c) transformation. According to Hughes (2005), replacement refers to integrating technology to replace other existing instructional methods and tools. Namely, at the replacement level, technology replaces other instructional *things* to serve the same instructional goal(s) without changing "instructional practices, student learning processes, or content goals" (p. 281). At the amplification level, technology increases the efficiency and effectiveness of task performance (Cuban, 1988; Pea, 1985 as cited in Hughes, 2005, p. 281), while transformation level also includes changing students' learning through "content, cognitive processes, and problem solving" (Pea, 1985, as cited in Hughes, 20025, p. 281) or teachers' instructional practices and roles" (Reinking, 1997, as cited in Hughes, 2005, p. 281).

## 4. METHODS

### 4.1. Research Design and Setting

This is a single case study (Yin, 2018) describing GenAI integration (i.e., Copilot powered by GPT-4) into an introductory undergraduate computer systems course. The course constituted an embedded single-case design (Yin, 2018) and was chosen because the course instructor intended to see how GenAI integration would work. Namely, the case satisfied Yin's (2018) *revelatory* and *longitudinal* rationales: It readily included GenAI integration that occurred throughout a semester. Case selection also aligned with Green's (2007) "pure gold sampling strategy" focusing on the most informative cases regarding the target entity that is initial intentional and exploratory GenAI integration in this study (p. 20).

The 12 teaching weeks included lecture and exercise sessions every week. Lectures occurred on Mondays from 10:15AM to 12PM while exercise sessions happened on Tuesdays from 8:15 to 10AM. The first lecture and exercise sessions provided an introduction into the course topics and introduced pros and cons of using GenAI. Further, the instructor-led lectures covered various topics ranging from boolean operations and gates to recursive structures. The teaching assistant-led exercise sessions covered solving coding and/or conceptual problems related to lecture topics in teams using personal computers. The students were allowed to use Copilot in their teams. Week 11 and week 12 exercise sessions were spared for Python programming in relation to a programming competition. Overall, the course content provided an intro into computer systems, encouraged students to achieve small-scale programming/coding, and asked students to practice their knowledge and skills using a programming language through face-to-face components and a learning management system.

### 4.2. Participants

Participants were a software engineering faculty and seven teaching assistants (TA). The course instructor (CI) had a doctoral degree in software engineering and had been faculty for 10 years as well as 25 years of professional engineering experience. Three TAs reported their descriptive info only: they had a mean age of 23 ($SD = 4.72$), and their years of experience as a TA was four months ($SD = 1.73$) on average. One TA had a B.A in software engineering and had six months of professional experience. Another one did not have professional experience and his highest degree was a high school diploma. Finally, the last TA had three months of professional engineering experience.

### 4.3. Instruments

#### 4.3.1. Semi-structured Interviews

To better understand the Copilot integration into the target course, we conducted multiple semi-structured interviews with the CI and TAs.

#### 4.3.2. Formative Practice Exercises

These exercises included multiple-choice questions covering factual and inference insights, and included three questions. Two extra exercises were run only once; so, they were not used.

### 4.4. Procedures

#### 4.4.1. GenAI Integration

After an initial introduction in the first week, Copilot was integrated into exercise sessions where students were supposed to work on practice problems, and lectures where CI discussed the use of the technology while coding. Students used Copilot mostly individually to solve the target problems and discuss the results in their teams in exercise sessions and lectures.

### 4.4.2. Data Collection

After Institutional Review Board (IRB) approval, two pilot interviews, one with a faculty, and one with a graduate student, were run to check how the interview protocols would work indicating no major issues. CI interviews were conducted at the beginning, in the middle and end of the semester, and the TA interview was conducted in the middle. Three interviews happened on Zoom, and one in Teams, and they were recorded and transcribed verbatim automatically and saved anonymously. Each interview took 40-60 minutes approximately.

### 4.4.3. Data Preparation

Quantitative data were checked for outliers, and the missing values in pre- and post-exercise scores were imputed using group median values. A sensitivity analysis comparing results from the imputed dataset and complete-case analysis revealed no statistically significant differences, supporting the imputation strategy. As for qualitative data, a native speaker who was not part of the team read through the transcripts and cleaned wording or highlighted unclear parts. The latter was clarified by the researchers by listening to the recordings, and the transcripts were revised for data analysis.

### 4.4.4. Data Analysis

Quantitative data were analyzed through R within RStudio (version 2025.05.0+.496) to examine any changes on students' formative practice exercise performance through pre- and post-exercises. Qualitative data were analyzed through thematic analysis (Braun & Clark, 2006):

familiarization, initial coding, theme development, reviewing themes, and defining and naming the themes. One researcher coded the whole data set and a second researcher double-coded them, and they came together to discuss all disagreements until consensus. The lowest initial agreement level was 86% for CI interviews, and 91% for the TA interview.

### 4.4.5. Trustworthiness

For credibility, qualitative data were coded by three researchers: two interview data sets were coded by two researchers, and the other two were coded by one of the researchers who also coded the first data set and a third researcher. The participants also provided members check insights (Creswell & Miller, 2000). Dependability is based on a thick description of the whole process thus enhancing transferability. Regarding confirmability, the researcher who coded the data initially took analytic notes or memos thus strengthening reflexivity.

#### 4.4.5.1. Positionality

The researchers have been approaching the current study with some pre-existing viewpoints. For instance, the first researcher has been doing research on educational technology, and thinks that technology integration should be based on how people learn better rather than focusing on tools only. The researcher also thinks that learners' prior knowledge is an important factor affecting further learning, and claims for increasing complexity and fidelity, and decreasing scaffolding over time. Namely, the researcher's approach to teaching and learning with technology includes both information processing-related and constructivist approaches depending on learners' knowledge level. The two other researchers, doctoral students, have similar approaches. The final three researchers have engineering backgrounds and value the role of hands-on experiential practice in EE.

## 5. RESULTS

GenAI Integration into Engineering Education

## 5.1. Descriptive Findings

Table 1 presents the descriptive findings:

| Exercise # | n | Median Pre | IQR Pre | Min. Pre | Max. Pre | Median Post | IQR Post | Min. Post | Max. Post |
|---|---|---|---|---|---|---|---|---|---|
| 1 | 114 | 1.27 | 1.11 | 0.65 | 2.8 | 2.05 | 1.00 | 0.4 | 3 |
| 2 | 96 | 1.25 | 1.00 | 0 | 3 | 2.62 | 1.06 | 0 | 3 |
| 3 | 85 | 1.00 | 1.00 | 0 | 3 | 1.00 | 1.00 | 0 | 3 |
| 4 | 69 | 1.12 | 1.00 | 0 | 3 | 1.75 | 1.50 | 0.25 | 3 |
| 5 | 57 | 2.00 | 0.83 | 0.5 | 3 | 2.33 | 1.17 | 0.33 | 3 |
| 6 | 85 | 1.50 | 1.00 | 0.25 | 3 | 2.00 | 1.00 | 0 | 3 |
| 7 | 44 | 2.00 | 1.00 | 0 | 3 | 3.00 | 1.00 | 1 | 3 |
| 8 | 39 | 1.25 | 0.75 | 0 | 2 | 1.75 | 0.62 | 0 | 3 |
| 9 | 39 | 2.00 | 0.66 | 0 | 3 | 2.00 | 1.00 | 0.33 | 3 |

Table 1. Descriptive Statistics for Pre- and Post-Tests
*Note.* Pre= Pre-test. IQR = Interquartile range. Min.= Minimum. Max. = Maximum. Post= Post-test

The Shapiro–Wilk test was used to evaluate the normality assumption. Results indicated violations of normality for Exercises 3, 4, 6, 7, and 8, thus leading to the use of Wilcoxon signed-rank test. Paired sample t-tests were applied to the remaining exercises where normality was met.

## 5.2. Changes in Students' Formative Practice Exercise Performance

Table 2 presents the results of the statistical tests.

Table 2. Gain Score Differences between Pre- and Post-Tests

| Exercise # | Method | Statistic | p | Mean Gain | SD Gain | Effect Size | Magnitude |
|---|---|---|---|---|---|---|---|
| | | | | | | | |

| 1 | Paired t-test | 9.93 | <.001 | 0.63 | 0.68 | 0.93* | large |
|---|---|---|---|---|---|---|---|
| 2 | Paired t-test | 10.08 | <.001 | 0.86 | 0.84 | 1.03* | large |
| 3 | Wilcoxon signed rank test | 966 | <.001 | 0.44 | 1.02 | 0.4 | medium |
| 4 | Wilcoxon signed rank test | 1396.5 | <.001 | 0.61 | 0.83 | 0.59 | large |
| 5 | Paired t-test | 2.53 | .014 | 0.25 | 0.73 | 0.33* | small |
| 6 | Wilcoxon signed rank test | 1579.5 | <.001 | 0.48 | 0.87 | 0.5 | large |
| 7 | Wilcoxon signed rank test | 217.5 | <.001 | 0.47 | 0.76 | 0.55 | large |
| 8 | Wilcoxon signed rank test | 439.5 | .001 | 0.49 | 0.85 | 0.53 | large |
| 9 | Paired t-test | 1.11 | 0.27 | 0.18 | 1.01 | 0.18* | negligible |

*Note.* * represents Cohen's d. All other effect sizes are Rosenthal's r.

Statistically significant improvements ($p < .05$) were found for all exercises except Exercise 9. As for effect sizes (Cohen's *d* and Rosenthal's *r*), large effects were observed in Exercises 1, 2, 4, 6, 7, and 8, indicating substantial improvement in post-exercise scores. However, medium and small effects were found for Exercises 3 and 5, respectively.

### 5.3. Interview Findings

#### 5.3.1. The First Interview with the Instructor

The first interview conducted with the CI in the first week of the semester revealed an overall ambivalent approach to GenAI integration and the following themes:

##### 5.3.1.1. Defining GenAI as Another Tool

The CI viewed GenAI as another tool at the beginning of the semester, which aligns with Hughes' (2005) replacement level of technology integration:

"So, this is just another tool. I used to use the books [and] photocopies in the past, you know. Then, there was the Internet. Google Search actually was AltaVista first, so we kind of advanced and now we have AI [and] we [will] probably [see] something else…"

### 5.3.1.2. Expected Benefits, Advantages and Usefulness

The CI expected GenAI to be useful from various perspectives including saving time, noting, "But if you really ask the right questions, you get a lot of information faster". Similarly, the CI assumes that GenAI can be a helpful and efficient tutor: "…if you don't know real time operating systems … you can actually, by having conversations [with AI], kind of learn about that much faster than reading several books…"

Moreover, the CI reported that GenAI can also be useful for writing references and quotations. The CI also indicated other general benefits of GenAI including easier translation among languages. Likewise, recording memories and making them accessible through GenAI is something that the CI found useful. Finally, the CI expressed that AI can help reduce anxiety by eliminating mistakes. These findings suggest that engineering instructors' technology integration would also relate to their technology use in other professional and general life tasks.

### 5.3.1.3. Healthy Skepticism and Disadvantages

The CI's initial insights included healthy skepticism about GenAI and its possible disadvantages, which included talking to others working in the private sector, noting: "I had this sort [of] chat about it from people from industry. And I think they are really confused: What to do and whether AI helps and not?". He also stated some distrust in relation to the credibility of GenAI outcomes and highlighted hallucination: "what I also found very very misleading is that Copilot gives references, typically at the bottom of the answer, for they found many times that the references have nothing to do with the answer".

The CI also voiced a disadvantage about learning the process of developing software in relation to GenAI's focus on quantity not quality: "The code quality is not really [related] to the amount of code. Probably it's [reverse/opposite], actually related to more code you have than you know the code quality is going to be lower…". The biggest disadvantage was lack of privacy and security though: "So yeah, can we use AI securely and safely? Yeah, we don't know the big players in the field. They are telling us that yes, you can use that and we are not using your data to learn about anything". This disadvantage is also related to compliance concerns the CI learned about through his business networks since he discovered that most local companies do not use GenAI due to compliance concerns. Consequently, the CI were actively reflecting on GenAI and its role not only in education but also in business regarding software engineering.

**5.3.1.4. Attitudes towards GenAI Policies in the Workplace**

The CI was surprised due to the lack of uniform GenAI policies around even though he liked the flexibility related to GenAI in their program and engineering, stating, "So, probably software engineering or engineering in general, we are kind of easygoing. That's my feeling, but all the areas are not". Interestingly, such a flexible atmosphere seems to have influenced the CI's attitudes towards AI positively: "So far it looks really good and is a new kind of new modern tool or technology".

**5.3.1.5. Potential Effects on and No Learning Trade-Off in Software Engineering**

The CI highlighted that GenAI hasn't yet had significant effects on software engineering despite some future possibilities: "we don't really see any impact yet because we have done and don't have enough experience. But it's expected that again it will make the development faster" by making repetitive tasks automatic. However, according to the CI there would still be an important no learning trade-off, stating, "It's the same thing you do over and over, but again you can write software faster, but my worry is that we are really developing a lot of software without really learning from it".

### 5.3.1.6. Practice-Oriented Engineering Education Pedagogy

The CI provided insights into the relationship between GenAI and EE pedagogy by focusing on both conceptual knowledge and practice and highlighted practice more. According to the CI, engineering is learned best through practice and learning from mistakes: "Maybe you can memorize something, but you actually learn by doing and actually trying to. [...] We learn by making mistakes, not just getting right answers, and I think that might be related to AI". Still**,** the CI pointed to the need for basic conceptual insights, but highlighted again that practice is essential: "…you actually have to try it yourself".

#### 5.3.1.6.1. Aspects of GenAI Integration into the Computing Course

The CI had strong positive expectations of GenAI and this stems from the CI's trying out the tool, and thought that GenAI helped students learn some basic information by providing the answers to questions, which increased students' engagement. The CI thinks that integrating GenAI would also encourage understanding:

"It's not only about learning how to do things, but also understanding code. So, Copilot allows you to highlight a piece of code and provide you with an explanation of it and then you can try that, you know, from different angles and you can analyze bigger chunks or smaller chunks of code and so on…".

The CI also used a book in his course and thought that it would still be useful for grasping conceptual content, highlighting, "I use a book. So, whenever the students have real issues with more theory questions or some kind of fundamentals, they will go [there]. And I mean that, I hope they can read things from the book". Likewise, the CI thought that teamwork was helpful in terms of GenAI integration, noting, "I'll also say that the teams will kind of help here…".

Importantly, the CI experimented with Copilot and checked how it could help with coding through Python. In this sense, the CI thought that it would be easy to find answers through GenAI, saying, "So, as for Python programming, you just go and search and you get like 1,000,000 hits when you have AI and GitHub Copilot integrated. You can always get the answers immediately to the problem that you have". Still, the CI highlighted that it is important to write codes: "We want to write code so again, uh, you are able to really, uh understand and learn pieces of code on your own…".

According to the CI, GenAI's answers may be speculative answers and lead to questionable learning outcomes. Consequently, the CI was not sure whether integrating GenAI would always lead to real learning. The CI still thought that students need to learn how to use GenAI in the course so that they can make the most out of them, which would serve the ultimate learning goal: "So the thing is that at the end, we want to, of course, we want, [to] make sure they learn Python and they can code something in Python".

**5.3.1.6.1.1. Conditions for Successful GenAI Integration into Engineering Education**

The CI emphasized that GenAI should not be doing everything so that students can learn how to code, claiming, "we shouldn't try to replace students' brains with AI because then actually then I would say we failed". Even though the CI finds teamwork valuable, they also prefer to employ individual work thinking that individual achievement would promote learning. Another condition was persuading students that they can learn engineering through GenAI, and that GenAI as a tutor would really work for software engineers. Learning by doing showed up again as a condition.

**5.3.2. The Second Interview with the Instructor**

The second interview with the CI was conducted in the middle of the semester producing the following themes:

**5.3.2.1. The Role of Prompts in Using GenAI Effectively**

The CI continued to explore other GenAI tools. For example, the CI highlighted that the quality of responses you get from ChatGPT depends largely on prompt quality. This quality also depends on the precision embedded into the prompts thus leading to more beneficial responses: "ChatGPT, the answers I got were really like completely off ... I actually started getting more and more precise but that basically means that my prompt was really getting precise…".

**5.3.2.2. Ambivalent Benefits, Advantages and Usefulness**

The CI reported that the students also started to use other GenAI tools in beneficial ways, stating, "I think what students say, I believe those are the good students, is that they use ChatGPT to confirm answers. They don't look for answers using AI tools. They use AI tools as a partner actually". Meanwhile, the CI also continued to explore GenAI use in business as it relates to coding. These insights showed that the CI was exploring GenAI integration in meaningful ways that mirror real-life situations: "So, they have Github Copilot licenses, all developers. But they really used that for our surprise to check some codes and then to generate documentation". Namely, the colleagues used GenAI to check their codes, but still checked the final outcomes of GenAI.

The CI kept an ambivalent attitude towards using GenAI to code since the outcomes may not be correct, highlighting, "It's not perfect but is a time saver; so, they still have to correct it". Interestingly, the ambivalent attitude becomes more significant when it comes to learning since the CI thought that learning needs effort and thinking, and GenAI can replace these factors:

"But, I think the challenge here is that with all the electronic tools also with the ChatGPT, Copilot because of the speed…You kind of try to optimize everything away, and that's why the learning also gets optimized, but we are real physical creatures…And actually, we need time to learn…"

Instead, the CI thought that students should use GenAI as a tutor: "what I encourage them to do is to use the AI tools as tutoring tools; so, as we know, like a way to have conversations and a way to kind of have a dialogue with a machine that knows things…" Accordingly the CI was happy with Copilot's feature that prevents giving direct answers and asks for dialogue.

### 5.3.2.3. Healthy Skepticism and Disadvantages

The CI indicated limited trust regarding GenAI tools in comparison with each other and it largely depended on the accuracy of the outcomes: "... ChatGPT just makes obvious mistakes that students from primary school can spot it". The CI also underscored GenAI's hallucination problem: "The answers they give are sometimes irrelevant... I just noted that they hallucinate…". So, by the middle of the semester, the CI's insights into the inaccuracy of GenAI-created outcomes were combined with the GenAI's potential to hallucinate. All these insights also led to some concerns about GenAI. It seems that students also made similar comparisons among tools and this affected their preferences: "So, we use Copilot. However, [a] common comment from students is that ChatGPT is better".

### 5.3.2.4. General and Course-Specific Concerns

The CI had some concerns about GenAI such as students' decreased engagement and decreased use of Copilot. The CI thought that the decreasing use could be due to other available resources that can be more advantageous: "...they have my notes and they know there's a

textbook…They don't have the incentives to use AI because they know they get precise data immediately by going in the PDF…".

The CI had general environmental concerns about GenAI that are ethical issues, noting, "Because we consume so much energy and water on all these in these data centers that is incredible…it's actually, they said it's really an environmental problem though". Finally, the CI also voiced concern about the possibility that coders would stop coding fully:

> "The risk here is however that we have engineers that kind of stop writing the code themselves…And this is confirmed by some engineers that after using Copilot they only tend to write very basic code or start writing and they wait for Copilot to finish…"

**5.3.2.5. Practice-Oriented Engineering Education Pedagogy**

The second interview results also emphasized learning by doing. Interestingly though, by the middle of the semester, lectures emerged together with practice:

> "... what students actually at least some students said is that the lectures are good but they really lack understanding because we have two hours of lectures and two hours of exercises. And they actually said that after the exercises are conducted, they understand the topics".

Learning by doing and practicing through exercises came with relevant feedback and insights from the CI and TAs: "...we let them solve those exercises and then we show them the answers afterwards. So, I explain it again when they are done with the exercise, I actually explain how that should be done". Still, the CI thought that textbooks, videos, and lectures are important for gaining the basics. After all, GenAI would not be helpful for gaining the basics: "But, if you read some part of the book or you just listen to me, you don't really have to use AI for that. Because those are basics; so, it's just basic understanding…"

**5.3.2.5.1  Aspects of GenAI Integration into the Computing Course**

The CI thought that Copilot was getting better without advertisements and started to retrieve previous conversations. The success of teamwork was also combined with students' expectations from GenAI, and members' contributions. Another factor impacting team success was students' level of readiness, which would make teams stronger: "...some groups are pretty strong. We have a lot of students that understand the topics or partly understand. Because we have students with a really different background from high school".

According to the CI, students shared their insights in their teams, and these conversations shaped their GenAI use through usefulness: "...my feeling is that what they do is more like they share the experience. And they will typically select tools that you know will use tools that work for them". Thus, the CI decided to model how to use GenAI while solving coding problems in class in a way that will evaluate the outcomes and provide feedback. The CI also preferred the TAs not to give answers directly but let the students first try to solve the problems by themselves, which aligns with the CI's thought that students should have strong insights into why they do what they do.

There were some negative aspects highlighted by the CI. For instance, GenAI was not useful when it came to more formal content including math. Similarly, the CI highlighted that the solutions needed in real-life business contexts would be more specific, and GenAI would not be very successful in that sense. The CI also highlighted that students' preference for other GenAI tools can affect GenAI integration, which would be linked to more advanced features of the preferred GenAI tools and students' familiarity with them: "I expected them to use Copilot, chat with them and a student took a mobile phone, [and] took the picture on the phone. And then, he started actually getting answer ... so that was actually the prompt for him…"

**5.3.2.5.1.1. Conditions for Successful Integration of GenAI into Engineering Education**

The CI highlighted some conditions that are necessary for successful GenAI integration. The first one related to students' learning and focused on class exercises. Then, the CI pointed to the importance of basic knowledge including quality prompts: "If you don't know the basics, you cannot really write the prompts. Because you have to instruct the tool for specific answers. So, you have to really narrow down". Having hardworking students in teams seems to have helped a lot too; but these students may not use GenAI and/or use different GenAI tools.

Another interesting condition was students' using it to ask for possible solution processes not solutions: "We should never ask for a solution. This is not correct. You should never ask to write this or solve this problem. This is wrong. You have to ask them about how they solved it". Likewise, the CI emphasized that students should also reflect on what they learn and do. Interestingly, the CI further connected the success of all these efforts with students' overall purpose and goal of pursuing a college degree highlighting that the students who just shoot for a diploma and are not genuinely interested in learning would not make the most of technological tools. The overall institutional approach also seems to affect instructors' GenAI integration, which seems to be driven by concerns about what students would need and do in business: "…we tend to be quite modern. So, we believe that first of all students in real life situations will use AI tools"

**5.3.2.6. The Future of GenAI in Coding/Programming**

The final theme that emerged out of the second interview with the CI was his future predictions about GenAI-enhanced coding. The CI emphasized that GenAI tools are not ready to replace human coders:

"The problem is that at the end, we have to deliver a program or application that solves specific problems that have certain characteristic features. And if AI can generate it, we don't need programmers for that. I'm sorry. That is the reality. But, I don't think the tools are really ready for that…".

**5.3.2.7. GenAI Integration based on Hughes' (2005) Three Levels**

In the second interview, the CI referred to the replacement and amplification levels of Hughes' (2005) technology integration model. Specifically, the CI indicated that GenAI can just replace existing tools and this shows a certain level of acceptance: "I think you can replace textbooks with generative AI... At least at the moment, I really think AI is just a tool and it's going to be like that so we have to kind of get used to it". The CI also indicated that GenAI can increase the efficiency and effectiveness of learning tasks without changing them.

**5.3.3. The Interview with Teaching Assistants**

**5.3.3.1. Defining GenAI as a Tool Mimicking Human Intelligence**

The TAs defined and agreed on GenAI as a technology that mimics human intelligence: "Maybe something that wants to mimic regular intelligence like humans or like behaviors of animals". They also related it to large language models: "When someone tells me about AI, it's a way to refer to language-based models".

**5.3.3.2. Ambivalent Benefits, Advantages and Usefulness**

The TAs highlighted the increased efficiency of completing tasks with GenAI. One said: "So, it definitely accelerates the research of any information. So, instead of doing [it] myself, I just typed an input into GPT and it printed everything it found for me online", and others agreed. Still, some others added that it is not completely positive: "Yes, it's useful to generate the structure of a website or some boilerplate, but, you cannot really create a whole application". These ways of using GenAI refer to Hughes's (2005) replacement and amplification levels.

### 5.3.3.3. A General Concern: GenAI Replacing Humans

A general concern raised by the TAs was that GenAI can replace humans: "So, some people will lose their jobs". Namely, TAs thought that GenAI would advance further and form an alternative for the human workforce. However, some TAs indicated that this replacement would not happen for software engineering and professions that entail creative thinking: "I don't say that software engineers will lose jobs. I don't. I just don't see it, but definitely it will exchange certain professions where creative thinking is not needed". This concern seemed to lead to negative attitudes towards GenAI as highlighted by one TA: "I just grow frustrated because every site I'm looking at it and getting suspicious".

### 5.3.3.4. Students' Attitudes towards GenAI

The TAs also commented on their students' attitudes toward GenAI. The first thing they noticed was that students were surprised that they were allowed to use GenAI in class: "…from what I've seen, the students were surprised that they could use AI in class". When explicitly asked whether it was positive or negative, they emphasized that it was just surprising: "I think it was neither. It was just surprising to them". Another TA added that the students were also excited: "…[they are] kind [of] excited about being able to use it and also that professor allows them and encourages them every exercise to use it in order to help them solve issues and find solutions".

### 5.3.3.5. General and Engineering-Specific Effects

The TAs highlighted GenAI effects including effects on life and engineering. The first one is about accessibility and the largest effect on engineering will be on solving engineering problems. However, some other TAs asserted that they do not see any big effects but small-scale ones currently: "Right now I just see it used and use it to write boilerplate code, because, more than that, I don't really see a big impact". Lastly, some TAs highlighted negative effects: "I think it would impact it for the worse".

### 5.3.3.6. Limitations of GenAI

The TAs expressed some GenAI limitations in relation to certain skills including creative thinking. Some even asserted that engineering is beyond GenAI and coding: "I believe engineering in general is way beyond coding, so I don't see an AI right now, I don't see an AI that would replace an engineer…". Accordingly, the TAs also pointed to the need for human involvement: "... even if, if the AI can actually do it, you always need to check if it's valid. So even if it can, the humans need to validate it; so, and for that, you need knowledge". Similarly, some TAs underscored double-checking GenAI-created outcomes: "every time the AI explains something, I need to double check it…".

### 5.3.3.7. Engineering Education Pedagogy Based on Practice and Theory

The TAs highlighted the importance of practice, but also indicated the need for theory. One said, "I mean, theory and practice have to meet somewhere, but, the practice part is definitely necessary for us…" while another one added, "I also think theory is important, but I agree with [name], that practice is probably the most important part". Still, another one pointed to keeping a balance between theory and practice: "I think it should be a balance between the theoretical part and applying it, be it through projects or similar".

When asked about GenAI's possible contributions to EE and its pedagogy, the TAs voiced ambivalent insights. In agreement with others, one noted: "I think it's good that we use it to know it's imitation. But again, besides some glorified search engine, I don't think it can substantially help". The TAs finally indicated that both theory and practice would be impacted by GenAI.

#### 5.3.3.7.1 Aspects of GenAI Integration into the Computing Course

The TAs expressed ambivalent and negative aspects of GenAI integration. For instance, focusing on the impossibility of making sure that students use it properly, one said and others agreed: "I believe it works here. It's working great, but there is always a thin layer of not knowing if the student is actually using it the right way".

Another negative aspect was hallucination: "So, it might go on 1% or another and you cannot stop hallucination. And we have seen very hilarious wrong results given by the AI even when explained step by step on exercise". One TA even highlighted that students can get frustrated: "...it's the small context window or like relatively small context window that the language models have, because the conversation stops after a while and they get really frustrated…".

##### 5.3.3.7.1.1. Conditions for Successful Integration of GenAI into Engineering Education

The TAs mentioned some conditions for a successful GenAI integration. For example, one TA asserted: "the students also have to understand that, if they get a prompt to AI that solves this exercise, they won't really learn from that". Likewise, another one emphasized students' understanding of coding problems and how to solve them. Accordingly, the TAs claimed that students should be critical GenAI users so that they can critically evaluate the outcomes.

The TAs also thought that teamwork can be helpful since it can encourage exchanging insights among students by comparing different GenAI-created outputs. An issue with teamwork observed by some TAs was that some team members would be more dominant. Then, the TAs switched to their suggestions about how to get students to make the most of GenAI. An interesting suggestion was providing an effective intro at the beginning of the semester including some direct instruction introducing students to how to use GenAI. Still, some other TAs thought that an introductory session would not be enough and it is necessary to provide sustainable help in the long term.

Lastly, the TAs reported that hardworking students who are important team success may not prefer to use GenAI and go for alternative traditional ways that would be also effective and efficient: "I never saw him using AI. He always does the research the traditional way.…So, he types the question and then he's looking on charts and forums for answers…". Finally, another TA referred to a more traditional alternative resource, the textbook: "What I've seen also by talking with them, they search it and they read all the book and that's where they get their information".

### 5.3.4. The Third Interview with The Instructor

#### 5.3.4.1. Benefits, Advantages and Usefulness

The CI highlighted that GenAI has potential as a translator in relation to coding for those who do not know multiple languages. The usefulness of GenAI as a tutor again showed up. This time, it seems that the students have already used GenAI as a tutor and positive outcomes emerged: "...we have a large number of students and students have specific questions about, for example, Python syntax. AI tools have no problem with that. So, getting help for learning programming is really nice… Students actually said that it's really great".

Still, based on Hughes' (2005) levels, we see a strong replacement of other instructional resources by GenAI while asking right questions, seeking assistance, learning, and investing enough time and effort. In this sense, the CI stated: "... General information is available, but you still need to spend time. Instead of talking to a person or reading a book, you talk to AI". Another advantage of GenAI seems to be its capability to categorize problems instead of directly solving them, which would help students solve the problems by themselves.

#### 5.3.4.2. Ambivalent Disadvantages and Relevant Concerns

The CI also pointed out some potential ambivalent disadvantages and corresponding concerns. The first thing was ineffectiveness regarding complex tasks, and the outcomes would be fun but incomplete: "I just wanted AI to compile a list of elements. I didn't remember something. It was actually a funny exercise and I noticed that the list was incomplete". Another interesting disadvantage was using GenAI for formal writing and reporting purposes even though it might be useful for format checking.

Due to such ambivalent disadvantages, the instructor continued to have limited trust: "AI is just kind of a fun tool to use. But it's not really truthful". This limited trust is also related to hallucination: "Some students just complained that the answers are nonsense". The final concern was privacy for the instructor: "The problem is with the privacy…".

#### 5.3.4.3. Engineering Education Pedagogy and Traditional Instruction

An interesting conversation also emerged on whether GenAI would turn education more traditional like writing codes and testing students' coding skills on paper in class. In this regard, the CI said that oral exams would become a viable option:

> "... the best probably exam is the oral exam. I believe this is the best way where you actually are one to one with the student and you can ask questions. Because, then, you can really see if the student has an understanding of certain topics or not"

The CI also mentioned talking to another professor working at a different university and learned that they were doing typical controlled computer-based exams where students couldn't use other online resources but had to focus on the exam itself. The professor even added that GenAI can add to and advance such closed systems.

### 5.3.4.3.1. Aspects of GenAI Integration into the Computing Course

The first aspect highlighted by the CI was a coding competition for teams that encouraged them to use GenAI. The CI suggested that such competitions can be done under controlled conditions. The CI also highlighted that the goal should be reflecting on coding deeply through specific problems, noting, "It's probably more realistic to ask, instead to solve like those very generic problems, to solve more specific problems where they actually need to be more creative". The CI also added that the problems should be authentic.

The CI kept a positive attitude towards teamwork, and this time, connected it to the observation that it is very common. The CI also repeated his thought that students should not use GenAI to get full answers or write the whole coding project. The CI came up with a way of detecting GenAI-produced projects based on the length of a project: "...if you really extensively use AI tools, you might have a problem because you might exceed the page limit and then you are in trouble".

### 5.3.4.3.1.1. Conditions for Successful Integration of GenAI into Engineering Education

The first condition was students' readiness again, and the CI emphasized that it is crucial to remember that learning takes time and effort: "Because learning involves effort and time, and students think about how to offload this. And I said, no, you cannot offload the learning part. This actually requires time and effort, and the tools help you in doing that.…"

Using well-defined coding problems in the course emerged as another condition, and the CI also suggested having a course covering GenAI and relevant basic knowledge.

Unsurprisingly, knowing about GenAI tools and how to use them was also important for the CI since the CI emphasized that some students may even not be able to send emails properly.

### 5.3.4.4. The Future of GenAI in Coding/Programming

The CI talked about the future of GenAI in coding stating that GenAI would change the nature of work and problems in software engineering. However, the CI also did not expect GenAI to replace coders completely since there would be a need to check and evaluate the product. Another reason for human involvement is that GenAI will not be able to solve all coding problems: "So, one thing we should demystify is that AI will solve all the problems. You still have to tell the tool what you need". All these points also relate to quality prompts, and the CI highlighted the importance of prompt engineering in the future linking it to knowledge: "Prompt engineering, right? It's it's kind of thing we have to learn".

## 6. DISCUSSION

This study investigated a software engineering CI's and TAs's initial intentional and exploratory experiences related to GenAI integration into an introductory undergraduate course. While earlier studies explored GenAI integration into EE from multiple perspectives, they have not substantially addressed initial GenAI integration efforts. Gaining insights into such a process can significantly inform and contextualize how engineering teaching staff approach and handle the integration of emerging technologies into their instruction. We found that an intentional and exploratory GenAI integration into EE would work well based on multiple factors including advantages, disadvantages, concerns, students' reactions, teamwork, and pedagogical approaches. Overall, the results provided significant insights into how GenAI can reshape teaching and learning practices in EE.

### 6.1. Formative Practice Exercise Performance

We found that GenAI integration can enhance students' performance on formative practice coding exercises thus aligning with research referring to positive GenAI effects (e.g., Birtek et al., 2023; Fan et al., 2023; Ho & Lee, 2023; Nikolic et al., 2023; Santos et al., 2024) despite other research referring to both limitations and challenges in EE (e.g., Ndiaye et al., 2023; Pham et al., 2023; Yelamarthi et al., 2024). However, the performance improvement may have also been affected by other factors that were not controlled. For instance, the novelty effect (e.g., Bjorkman et al., 2019; Oppmann et al., 2025; Shin et al., 2019) that occurs when people start to use a new technology for the first time thus leading to increased use, which would weaken over time, may have created a short-term excitement and increased students' performance. Likewise, the CI's and TAs's increased efforts on learning about GenAI integration could have helped students enhance their performance. Still, increased post-exercise performance after GenAI-enhanced instruction indicates a connection between the two.

**6.2. Intentional and Exploratory GenAI Integration**

Our findings indicated that intentional and exploratory experiences can form an important way of GenAI integration into EE. Importantly, the participating CI and TAs eagerly tried to integrate GenAI into their engineering course, questioning how to do it meaningfully and actively trying to solve issues that may have shown up. These findings align with earlier technology integration studies focusing on teachers' approaching new technology with skepticism and evaluating its contributions (e.g., Albaugh, 1997), and claims that faculty should take the initiative to deal with GenAI integration even without institutional support (e.g., Hassani et al., 2025). This is a form of healthy skepticism leading to further evaluation and investigation since it questions whether a technology would help teach better (Albaugh, 1997). However, faculty's professional tendency to be sceptical and critical can also impede technology integration (e.g., Tabata & Johnsrud, 2008).

The findings also align with the earlier finding that faculty's first-hand experiences with an innovation can create positive attitudes that lead to more adoption (e.g., Tabata & Johnsrud, 2008). The healthy skepticism and further evaluation can also shape engineering instructors' beliefs (e.g., Ertmer, 1999; Ertmer et al., 2012) about the value of GenAI. The beliefs or internal barriers are generally stronger than external barriers ranging from resources to institutional support (Ertmer, 1999; Miranda & Russell, 2012). In this study, the CI and TAs were evaluating GenAI use and forming their beliefs and attitudes in relation to its function. When teachers have positive attitudes and beliefs, they can overcome the external barriers (Ertmer et al., 2012; Ertmer & Ottenbreit-Leftwich, 2010), and such positive attitudes can also encourage faculty's involvement in technology integration (Tabata & Johnsrud, 2008).

The current findings referring to the CI's and TAs's ongoing evaluation of GenAI integration also aligns with reflective practice or teaching (e.g., Schon, 2017; Baporikar, 2016; Brinegar, 2023; Machost & Stains, 2023), teacher professional development (e.g., Korkko et al., 2016; Meyer et al., 2023) and faculty professional development (e.g., Efu, 2023; Hora & Smolarek, 2018; Kirpalani, 2017). For example, both the CI and TAs implemented reflection for action, reflection in action and reflection on action while attempting to integrate GenAI. Namely, the participants employed reflection before integrating GenAI, spontaneous reflection on what is working or not, and delayed reflection that focused on what happened and whether it was working. Efu (2023) highlighted that reflection is an integral part of continuous professional development for faculty. Such critical reflections also align with deliberate practice that encourages expertise through targeted and repeated practice, and immediate feedback (e.g.,

Ericsson et al., 1993; Ericsson & Harwell, 2019), and the positive relationship between deliberate practice and enhanced teaching performance (e.g., Ellison & Woods, 2016; Marcus et al., 2020; Rowat et al., 2021). To this end, the CI and TAs employed active reflections on GenAI integration and got feedback from students and each other.

  Through all those reflections, the CI's perspectives on GenAI integration evolved: At the beginning, the CI had positive expectations that moved towards more ambivalent expectations by the middle of the semester, and mainly stayed so towards the end. Given that it was the first course into which the CI was trying to integrate GenAI, going from high to ambivalent expectations is understandable since they would have become more realistic over time. This trend aligns with both diffusion of innovations (e.g., Rogers, 1987; Dearing & Cox, 2018) that predicts ambivalence between early adopters and early majority evaluating relative advantages, and technology acceptance model (e.g., Davis, 1989; Davis & Venkatesh, 1996; Shin et al., 2022) that expects high expectations to weaken through lower perceived usefulness. After all, the CI and TAs had some skepticism, concerns due to some ambivalent advantages and benefits.

  Accordingly, it is reasonable to state that the participating CI and TAs were in the early adoption stage. In other words, the GenAI integration into EE in this study aligns with existing theoretical and conceptual expectations, and does not represent an abrupt change. The continuous intentional and exploratory GenAI integration also indicates that the CI and TAs were motivated enough to learn about this process, which aligns with Ibrahim and Nat's (2019) point that instructor motivation can promote the adoption of new technologies. Namely, the CI and TAs did not have strong resistance against GenAI, which is a common issue (Moser, 2007; Renes & Strange, 2011). These insights also fit into the main themes identified, and how they related to each other (Table 3).

GenAI Integration into Engineering Education

Table 3. Matching the Main Themes Across Interviews

| Instructor Interviews | | | TA Interview |
|---|---|---|---|
| 1st Interview | 2nd interview | 3rd Interview | |
| > Defining GenAI as Another Tool | > The Role of Prompts in Using GenAI Effectively | | > Defining GenAI as a Tool Mimicking Human Intelligence |
| > Expected Benefits, Advantages and Usefulness | > Ambivalent Benefits, Advantages and Usefulness | > Benefits, Advantages and Usefulness & Ambivalent Disadvantages and Relevant Concerns | > Ambivalent Benefits, Advantages and Usefulness |
| > Healthy Skepticism and Disadvantages | > Skepticism and Disadvantages<br>> General and Course-Specific Concerns | | > A General Concern: GenAI Replacing Humans & Limitations of AI<br><br>> Students' attitudes toward GenAI |
| > Attitudes toward GenAI Policies/ Regulations in the Workplace | | | |
| > Potential Effects on and No Learning Trade-Off in Software Engineering | | | > General and Engineering-Specific Effects<br><br>> Limitations of GenAI |
| > Practice-Oriented Engineering Education Pedagogy | > Practice-Oriented Engineering Education Pedagogy | > Engineering Education Pedagogy and Traditional Instructional Elements | > Engineering Education Pedagogy |
| | > The Future of GenAI in Coding/Programming | > The Future of GenAI in Coding/Programming | |
| > Hughes (2005): | > Hughes (2005): | > Hughes (2005): | > Hughes (2005): |

| Replacement | Replacement and Amplification | Replacement | Replacement and amplification |

The main themes were quite similar across the interviews, and by the end of the semester, the CI still seems to be in an ambivalent mode regarding GenAI integration, and the TAs had ambivalent attitudes in the middle of the semester. The CI's ambivalent attitudes at the end of semester may suggest that one semester would not be enough to make more conclusive decisions. Given that other CIs and TAs can also have similar ambivalent attitudes, reflecting on them by interacting and sharing insights with each other can help. This is in line with previous research supporting faculty professional development through peer interactions (e.g., Agbonlahor, 2006; Nicolle & Lou, 2008). For instance, when faculty show the relative advantages of a technology for themselves and others, the adoption of technologies can be enhanced (Agbonlahor, 2006).

After all, when insights are not forced but come from colleagues, adoption can be more effective (Nicolle & Lou, 2008). One caveat regarding such collaborative faculty technology integration would be possible individual differences among academics such as self-efficacy (Kulviwat et al., 2014), their roles (Gilbert & Kelly, 2005), concerns including risk avoidance (Lin & Cantoni, 2018), and incorrect information and privacy (Menekse, 2023). Therefore, it is reasonable to take all such potential individual differences into account while planning any faculty professional development regarding emerging technology integration.

## 6.3. GenAI Integration and Hughes's (2005) Technology Integration Levels

Technology integration is not only about technologies but also processes: achieving instructional goals (Hew & Brush, 2007), responding to in-class situations and issues (Roblyer & Doering, 2010), and running informed practices to enhance student learning (Kim et al., 2013).

So, technology integration should be taken as a process that does not just replace old teaching ways (Kim et al., 2013). This study's finding that GenAI integration stayed at replacement and amplification levels, and did not turn into transformation is understandable since it was the first time for the CI and TAs trying to integrate GenAI. It also aligns with the early adoption phase of GenAI where questioning its relative advantages and disadvantages continues. Interestingly, amplification finding also indicated that GenAI did not only replace some older instructional elements but also increased efficiency. One concrete example turned out to be fast-paced searching for relevant information. All these insights align with previous research suggesting that trying out new technologies by evaluating consequences can be useful (e.g., Chan et al., 2016).

Lastly, even though replacement and amplification levels are also in line with real-life GenAI use in software engineering thus increasing automation, transformation is still possible since GenAI seems to transform coding processes. For instance, transformational GenAI integration can work through creativity, problem solving and ideation (Jackson et al., 2024). Other research also showed that GenAI can enhance code generation and test optimization but still needs human involvement (e.g., Poldtrack et al., 2023). Similarly, Denny et al. (2026) claimed that Github Copilot can be successful at solving some programming problems and human prompt engineering can increase its capacity. These insights refer to GenAI-human collaboration or a more hybrid approach including humans and GenAI (e.g., Fan et al., 2023; Sengul et al., 2024). However, as GenAI's capabilities in software development continue to advance, direct human involvement in coding and implementation activities may progressively diminish, with human expertise increasingly focusing on high-level conceptual design, system architecture, and strategic software planning.

**6.4. Implications**

The current findings refer to the importance of understanding engineering teaching staff's experiences of emerging technology integration since they inform how to integrate technology into EE (e.g., Naser, 2022). This way, informed guidance can be formed and shared with engineering faculty, which is important for accreditation purposes (Naser, 2022). Likewise, given that innovation adoption can be facilitated by trying out (Abrahams, 2010), faculty can be encouraged to experiment with new technologies. Such first-hand experimenting can strongly lead to positive attitudes, which can further enhance adoption (Tabata & Johnsrud, 2008). To this end, considering the advantages, disadvantages, concerns, and engineering teaching staff's pedagogical approaches is important for institutions to encourage effective GenAI integration.

Reflective and deliberate practice through communities of practice based on engineering teaching staff's intentional GenAI integration efforts can help integrate GenAI in more informed ways, which aligns with the benefits of similar knowledge sharing (e.g., Latif, 2017). This can also be a collaborative faculty professional development in practice, and institutions can encourage it (Hora & Smolarek, 2018) especially when little is known about an emerging technology. After all, interactions with other faculty and collegiality are directly linked to faculty's learning and adopting innovations (Nicolle & Lou, 2008), and can enhance overall technology adoption (Abraham, 2010) through learning by doing (Al Badi et al., 2024). Based on these efforts, institutions can prepare institutional GenAI integration guidelines since early faculty adopters' experimenting with GenAI can inform future policies (Hassani et al., 2025). The current research also suggests that TAs or other teaching staff can also contribute.

**6.5. Limitations and Suggestions for Further Research**

The results should be carefully approached due to some limitations. First, this study used a single case approach, and analyzed it deeply from multiple perspectives to present more comprehensive descriptions. Namely, we used one single introductory computer systems course in a specific semester and institutional context, which refers to what Kim et al. (2013) calls a microcontext. Therefore, we need more research conducted in different engineering course contexts including other institutions (Kurtz et al., 2024) since broader generalization requires multi-year, and multi-site evaluations (Tahiri et al., 2023). Second, it is also possible to design and develop learning experiences in engineering courses by using GenAI (e.g., Ho & Lee, 2023), which did not emerge in this study. Thus, future research can also look at instructors' and other teaching staff's GenAI-enhanced content design and development experiences.

Furthermore, GenAI models with general purposes can underperform in specialized EE contexts. Consequently, research can also focus on developing domain-specific GenAI trained on tailored datasets related to engineering tasks. These targeted tools can produce significant enhancements in educational outcomes, providing precise and relevant support for complex engineering tasks. Another limitation is that a CI and seven TAs were participants only. Future research examining students' perspectives can provide complementary insights since students also use, do not use (Dai, 2025; Johri et al., 2024; Li et al., 2025), or use GenAI more than their instructors (Guillen-Yparrea & Hernandez-Rodríguez, 2024). Likewise, engineering students' needs (Kozan et al., 2021) and study strategies (Butt et al., 2021) that relate to their academic success can also affect GenAI integration.

Finally, GenAI students' GenAI use was mainly limited to solving coding problems. Thus, we need more research on GenAI integration regarding different types of student experiences. For instance, engineering students can use GenAI for design and development purposes. Likewise, GenAI can also be used to develop effective instructional content including various examples and explanations (Mollick & Mollick, 2023). Such attempts can also enhance the status of technology in STEM education since technology has been integrated into STEM education less compared to science and engineering (Kozan et al., 2023).

## 7. CONCLUSIONS

This study investigated how a software engineering CI's and TAs's first-hand intentional and exploratory GenAI integration endeavors in an introductory undergraduate course would work, and how these would affect student performance on formative exercises. We found that engineering teaching staff's initial experimental experiences of GenAI integration is multidimensional and integration can stay at the levels of replacing other instructional elements and increasing efficiency. Still, reflective and critical experimental GenAI integration experiences can be useful and provide important insights into what would work. Such experiences can also lead to more realistic expectations and encourage teaching staff to explore more thus turning the whole process into a collaborative professional development. We also found that GenAI integration can enhance students' formative exercise performance, and they would be surprised and excited about using GenAI even though they may use it less over time. Consequently, higher education institutions can support engineering teaching staff's experimental technology integration, and use such experiences to keep pace with technological innovations.

## 8. Statement on Artificial Intelligence

Artificial intelligence was used for copyediting and proofreading purposes.

GenAI Integration into Engineering Education

GenAI Integration into Engineering Education

GenAI Integration into Engineering Education

GenAI Integration into Engineering Education

GenAI Integration into Engineering EducationNavigating challenges and leveraging opportunities. *Sustainability*, *17*(7), 3201. https://doi.org/10.3390/su17073201

Latif, F. (2017). TELFest: an approach to encouraging the adoption of educational technologies. *Research in Learning Technology, 25,* 1-14. 10.25304/rlt.v25.1869

Lee, C. C., & Low, M. Y. H. (2024). Using genAI in education: The case for critical thinking. *Frontiers in Artificial Intelligence*, *7*, 1452131.

Lesage, J., Brennan, R., Eaton, S. E., Moya, B., McDermott, B., Wiens, J., & Herrero, K. (2024). Exploring natural language processing in mechanical engineering education: Implications for academic integrity. *International Journal of Mechanical Engineering Education*, *52*(1), 88–105. https://doi.org/10.1177/03064190231166665

Li, R., Li, M., & Qiao, W. (2025). Engineering students' use of large language model tools: An empirical study based on a survey of students from 12 universities. *Education Sciences, 15*(3), 280. https://doi.org/10.3390/educsci15030280

Li, Y., Ji., W., Liu, J., & Li, W. (2024). Application of generative artificial intelligence technology in customized learning path design: A new strategy for higher education. *Proceedings of International Conference on Interactive Intelligent Systems and Techniques (IIST),* IEEE. 10.1109/IIST62526.2024.00099

Lin, J., & Cantoni, L. (2018). Decision, implementation, and confirmation: Experiences of instructors behind tourism and hospitality MOOCs. *International Review of Research in Open and Distance Learning, 19*(1), 275–293. https://doi.org/10.19173/irrodl.v19i1.3402

Liu, C., & Yang, S. (2024). Application of large language models in engineering education: A case study of system modeling and simulation courses. *International Journal of

GenAI Integration into Engineering Education

GenAI Integration into Engineering EducationMoser FZ (2007) Faculty adoption of educational technology. *EDUCAUSE Quarterly 1,* 66-69.

https://er.educause.edu/~/media/files/article-downloads/eqm07111.pdfMtebe

Mozgovoy, M., Burelli, P., Sánchez-Rada, L. M., & others. (2023). ChatGPT Challenges Blended Learning Methodologies: A Case Study in Programming Education. *Applied Sciences*.

Murali, R., Ravi, N., & Surendran, A . (2024). Augmenting virtual labs with artificial intelligence for hybrid learning. In *2024 IEEE Global Engineering Education Conference (EDUCON)*, 1–10. 10.1109/EDUCON60312.2024.10578649

Mustapha, K. B., Yap, E. H., & Abakr, Y. A. (2024). Bard, ChatGPT and 3DGPT: A scientometric analysis of generative AI tools and assessment of implications for mechanical engineering education. *Interactive Technology and Smart Education*, *21*(4), 588–624. https://doi.org/10.1108/ITSE-10-2023-0198

Naser, M. Z. (2022). A Faculty's perspective on infusing artificial intelligence into civil engineering education. *Journal of Civil Engineering Education 148*(4), https://doi.org/10.1061/(ASCE)EI.2643-9115.0000065

Nicolle, P. S., & Lou, Y. (2008). Technology adoption into teaching and learning by mainstream university faculty: A mixed methodology study revealing the "how, when, why, and why not". *Journal of Educational Computing Research, 39*(3), 235-265. https://doi.org/10.2190/EC.39.

Ndiaye, Y., Lim, K. H., & Blessing, L. (2023). Eye tracking and artificial intelligence for competency assessment in engineering education: a review. *Frontiers in Education, 8,* 1170348. https://doi.org/10.3389/feduc.2023.1170348